\documentclass[submission,copyright,creativecommons]{eptcs}
\providecommand{\event}{LSFA 2012} 
\usepackage{graphicx}
\usepackage{moreverb}
\usepackage{cite,xcolor}
\usepackage{amsmath,amssymb,amsfonts}
\usepackage{listings}
\usepackage[all]{xy}
\usepackage{xcolor}
\usepackage{verbatim}
\usepackage{multirow}

\def\titlerunning{On Formalizing Confluence of Orthogonal Systems}
\def\authorrunning{A.C.R. Oliveira \& M. Ayala-Rinc\'on}

\begin{document}

\title{Formalizing the Confluence of Orthogonal Rewriting
  Systems\thanks{Work supported by grants from CNPq/Universal, CAPES/STIC-AmSud and
     FAPDF/PRONEX.}}

\author{Ana Cristina Rocha Oliveira\thanks{Author supported by CAPES.}   \; and \, Mauricio Ayala-Rinc\'{o}n\thanks{Author partially supported by CNPq.}\\
  Grupo de Teoria da Computa\c{c}\~{a}o\\
  Departamentos de Matem\'atica e Ci\^encia da Computa\c{c}\~ao\\
  Universidade de Bras\'{i}lia \\
  Bras\'{i}lia D.F., Brazil\\
  Email: {\tt anacrismarie@gmail.com, ayala@unb.br}}

\date{}

\maketitle

\begin{abstract}
  Orthogonality is a discipline of programming that in a syntactic
  manner guarantees determinism of functional 
  specifications. Essentially, orthogonality avoids, on the one side,
  the inherent ambiguity of non determinism, prohibiting the existence
  of different rules that specify the same function and that may apply
  simultaneously (\emph{non-ambiguity}), and, on the other side, it
  eliminates the possibility of occurrence of repetitions of variables
  in the left-hand side of these rules (\emph{left linearity}). In the
  theory of term rewriting systems (TRSs) determinism is captured by
  the well-known property of confluence, that basically states that
  whenever different computations or simplifications from a term are
  possible, the computed answers should coincide. Although the proofs
  are technically elaborated, confluence is well-known to be a
  consequence of orthogonality. Thus, orthogonality is an important
  mathematical discipline intrinsic to the specification of recursive
  functions that is naturally applied in functional programming and
  specification.  Starting from a formalization of the theory of TRSs
  in the proof assistant PVS, this work describes how confluence of
  orthogonal TRSs has been formalized, based on axiomatizations of
  properties of rules, positions and substitutions involved in
  parallel steps of reduction, in this proof assistant. Proofs for
  some similar but restricted properties such as the property of
  confluence of non-ambiguous and (left and right) linear TRSs have
  been fully formalized.
  
\end{abstract}

\section{Introduction}

Termination and confluence of term rewriting systems (TRSs) are
well-known undecidable properties that are related with termination of
computer programs and determinism of their outputs.  Under the
hypothesis of termination, confluence is guaranteed by the critical
pair criterion of Knuth-Bendix(-Huet) \cite{KnBe70, Hu80},
which establishes that whenever all critical pairs of a given
terminating rewriting system are joinable, the system is
confluent. This criterion as well as other criteria for abstract reduction systems
such as Newman's lemma were fully formalized in the proof assistant
PVS in \cite{GaAR2010, GaAR2008c} over the PVS \emph{theory} {\tt trs}
\cite{GaAR2008b}, that is available in the NASA LaRC PVS library
\cite{NASAPVSLib}.  Without termination, confluence analysis results
more complex, but several programming disciplines, from which one
could remark orthogonality, guarantee confluence without the necessity
of termination.

In the context of the theory of recursive functions and functional
programming as in the one of TRSs, the programming discipline of
orthogonality follows two restrictions: \emph{left-linearity} and
\emph{non-ambiguity}.  The former restriction allows only definitions
or rules in which each variable may appear only once on the left-hand
side ({\bf lhs}, for short) of each rule; the latter restriction
avoids the inclusions of definitions or rules that could
simultaneously apply.

This work reports a formalization of the property of confluence
of orthogonal systems in the proof assistant PVS. The formalization
uses the PVS \emph{theory} {\tt trs}, but several additional notions
such as the one of parallel rewriting relation were included in order
to follow the standard inductive proof approach of this property, that
is based on the proof of the diamond property for the parallel
reduction associated to any orthogonal TRS as presented in
\cite{BaNi98}.  In the current state of this formalization, several
technical details that are related with properties of terms and
subterms involved in one-step of parallel reduction are
axiomatized. Additionally, the PVS \emph{theory} includes a complete
formalization of the confluence of non-ambiguous and linear TRSs.  An
extended version of this paper as well as the PVS development are
available at {\tt www.mat.unb.br/$\sim$ayala/publications.html}.

Proofs of confluence of orthogonal TRSs have been known at least since
Rosen's seminal study on Church-Rosser properties of \emph{tree
  manipulating systems} \cite{Ro1973} and several of them are based on
a similar strategy through the famous Parallel Moves Lemma.  Rosen's
proof uses a notion of residuals of positions in a notation that was
standardized further by Huet in \cite{Hu80}, paper in which Huet
presented a proof of confluence of left-linear and parallel closed
TRSs that unlike Rosen's proof admits critical pairs that should be
joinable from left to right in a sole step of parallel reduction.

In the chapter on orthogonality of \cite{Te2003}, the authors
presented five styles of proof of confluence of orthogonal systems as
well as an extension of the confluence result to weakly orthogonal
TRSs. All the given styles of proof are not different in essence: the
first one uses the notions of residuals and descendants via the
parallel moves lemma, the second one avoids explicit mention of
residuals by underlining reductions, the third one imports from
$\lambda$-calculus and combinatory logic the notion of complete
developments and the fourth style uses elementary and reduction
diagrams.  The fifth given proof is an inductive confluence proof that
is the more related with our approach of formalization and follows
lines of reasoning based on analysis of properties of the parallel
rewriting relation and the parallel moves lemma, just by changing the
definition of parallel reduction. In this proof the parallel relation
is defined from the rewriting relation as the reflexive relation that
is compatible with substitutions and, parallely  compatible with
contexts. Thus, after proving a version of the parallel moves lemma,
the diamond property of the parallel reduction is proved by induction
on the structure of terms based on the analysis of the six possible
cases of a parallel divergence; that is, whether the divergence terms
are obtained from a term by application of two different steps of
parallel reduction, by combinations of reflexivity, substitution and
context according to the definition of the parallel relation.  In this
analysis, the version of the parallel moves lemma is applied for the
case of a divergence in which on the one side a term is obtained by
substitution and on the other side by context.

For this formalization it has been chosen the inductive proof presented in
\cite{BaNi98} because it uses the nowadays standard rewriting notation
as the PVS \emph{theory} {\tt trs} does,  uses a standard
definition of parallel reduction and follows lines of
reasoning that from the authors'  viewpoint are of great didactical interest.

\section{Specification of basic Notions and Definitions}

Standard notation of of the theory of rewriting is used as in
\cite{BaNi98} or \cite{Te2003}. One says that a rewriting relation
$\rightarrow$: 

\begin{center}\begin{tabular}{l@{\hspace{2cm}}l}
is \emph{confluent} whenever
&$({}^*\!\!\leftarrow \circ \rightarrow^*)\subseteq(\rightarrow^* \circ \;{}^*\!\!\leftarrow),$ \\[1mm]
\emph{triangle-joinable} if &
$(\leftarrow \circ \rightarrow)\subseteq (\rightarrow \circ \;{}^=\!\!\leftarrow)
\cup(\rightarrow^= \circ\;\leftarrow),$\\[1mm] 
has
the \emph{diamond property} if
&$(\leftarrow \circ \rightarrow)\subseteq(\rightarrow \circ
\leftarrow).$\\[1mm]
\end{tabular}\end{center}

A well-defined set of terms is built from a given signature and an enumerable
set of variables. A rule $e = (l,r)$ is an ordered pair of terms such
that the first one cannot be a variable and all variables in the
second one occur in the first one. A TRS is given as a set of rules.
The reduction relation $\rightarrow_E$ induced by a TRS $E$ is built
as follows: a term $t$ reduces to $t_0$ (denoted as $t\rightarrow
t_0$) if there are a position $\pi$ of $t$, a rule $e\in
E$ and a substitution $\sigma$ such that: $t|_{\pi}\:=\:lhs(e)\sigma$,
i.e., the subterm of $t$ at position $\pi$ is the lhs of the rule $e$
instantiated by the substitution $\sigma$; and $t_0$ is obtained from
$t$, by replacing the subterm at position $\pi$ by the corresponding
instantiation of the right-hand side ({\bf  rhs}, for short)  of the rule, 
that is $rhs(e)\sigma$.  The only change done
in order to obtain $t_0$ from $t$, occurs at position $\pi$.  All
this is summarized by the following notation: $t \;\;= \;\;
t[\pi\leftarrow lhs(e)\sigma]\;\;\;\;\rightarrow_E\;\;\;\;
t[\pi\leftarrow rhs(e)\sigma]\;\;=\;\; t_0$, where, in general,
$u[\pi\leftarrow v]$ denotes the term obtained from $u$ by replacing
the subterm at position $\pi$ of $u$ by the term $v$.

Given terms $t_1$ and $t_2$, one says that $t_1$ reduces in parallel
to $t_2$, denoted as $t_1\:\rightrightarrows_E\:t_2$, whenever there
exist finite sequences
$$\Pi :=\pi_1, \ldots, \pi_n;$$ 
$$\Sigma:=\sigma_1,\ldots,\sigma_n\mbox{ and}$$  
$$\Gamma := e_1,\ldots, e_n$$ 
of parallel positions of $t_1$, substitutions and rules in $E$,
respectively, such that:
$$t_1|_{\pi_i}\:=\:lhs(e_i)\sigma_i,\;\forall i=1,\dots,n,$$ 
i.e., the subterm of $t_1$ at position $\pi_i$ is the lhs of the rule
$e_i$ instantiated by the substitution $\sigma_i$; and $t_2$ is
obtained from $t_1$, by replacing all subterms at positions in $\Pi$
as
$$t_2|_{\pi_i}\:=\:rhs(e_i)\sigma_i,\;\forall i=1,\dots,n,$$ 
i.e., for all $i$, the subterm at position $\pi_i$, that is the
$\sigma_i$ instance of the lhs of the rule $e_i$, $lhs(e_i)\sigma_i$,
is replaced by the $\sigma_i$ instance of the rhs of the rule,
$rhs(e_i)\sigma_i$.  The only changes done in order to obtain $t_2$
from $t_1$, occur at the positions in $\Pi$.  All this is summarized
by the following notation:
$$t_1 \;\;= \;\;t_1[\pi_1\leftarrow l_1\sigma_1]\ldots[\pi_n\leftarrow l_n\sigma_n]\;\;\;\; \rightrightarrows_E \;\;\;\;t_1[\pi_1\leftarrow r_1\sigma_1]\ldots[\pi_n\leftarrow r_n\sigma_n] \;\;=\;\; t_2,$$
where, $l_i = lhs(e_i)$ and $r_i = rhs(e_i)$, for $1\leq i \leq n$.

The PVS \emph{theory} {\tt trs} includes all necessary basic notions
and properties to formalize elaborated theorems of the theory of
rewriting such as the one of confluence of orthogonal systems.  {\tt
  trs} includes specifications and formalizations of the algebra of
terms, subterms and positions, properties of abstract reduction
systems, confluence and termination, among others.  The current
development of the PVS \emph{theory} called {\tt orthogonality} deals
specifically with orthogonality related notions and properties. Among
the definitions specified inside the \emph{theory} {\tt orthogonality}
one could mention the basic boolean ones listed below, where {\tt E}
is a set of rewriting rules (equations).

{\footnotesize
\begin{verbatim}
- Ambiguous?(E): bool =  EXISTS (t1, t2) : CP?(E)(t1,t2)

- linear?(t): bool = FORALL (x | member(x,Vars(t))) : Card[position](Pos_var(t,x)) = 1 

- Right_Linear?(E): bool = FORALL (e1 | member(e1, E)) : linear?(rhs(e1))   

- Left_Linear?(E): bool = FORALL (e1 | member(e1, E)) : linear?(lhs(e1))   

- Linear?(E): bool = Left_Linear?(E) & Right_Linear?(E)

- Orthogonal?(E): bool =  Left_Linear?(E) & NOT Ambiguous?(E)
\end{verbatim}
}

In the specification of {\tt Ambiguous?(E)}, {\tt CP?(E)(t1,t2)}
specifies that {\tt t1} and {\tt t2} are critical pairs of the
rewriting system {\tt E}. A term {\tt t} is {\tt linear?} whenever,
each variable {\tt x} in {\tt t} occurs only once. The expressions
{\tt Right\_Linear?(E)} and {\tt Left\_Linear?(E)} indicate
respectively that the rhs and the lhs of all rules in {\tt E} are
linear. The predicate {\tt Linear?} specifies linearity of sets of
rewriting rules. Finally, {\tt Orthogonal?}  specifies orthogonality
of TRSs.

More elaborated auxiliary definitions are specified as:

{\footnotesize
\begin{verbatim}
- local_joinability_triangle?(R) : bool = FORALL(t, t1, t2) : R(t, t1) & R(t, t2) =>
             EXISTS s : (RC(R)(t1, s) & R(t2, s)) OR (R(t1, s) & RC(R)(t2, s))

- replaceTerm(s: term, t: term, (p: positions?(s))): RECURSIVE term =
     IF length(p) = 0  THEN t
     ELSE LET st = args(s), i = first(p), q = rest(p),  
                        rst = replace(replaceTerm(st(i-1), t, q), st,i-1) IN app(f(s), rst)
     ENDIF MEASURE length(p)

- reduction?(E)(s,t): bool =  EXISTS ( (e | member(e, E)), sigma, (p: positions?(s))): 
      subtermOF(s, p) = ext(sigma)(lhs(e))  &  t = replaceTerm(s, ext(sigma)(rhs(e)), p)

- replace_par_pos(s, (fsp : SPP(s)), fse | fse'length = fsp'length, 
                          fss  | fss'length = fsp'length)  RECURSIVE term =
     IF length(fsp) = 0 THEN s  
     ELSE replace_par_pos(replaceTerm(s, ext(fss(0))(rhs(fse(0))), 
                                                fsp(0)), rest(fsp), rest(fse), rest(fss))
     ENDIF MEASURE length(fsp)

- parallel_reduction?(E)(s,t): bool =  
   EXISTS (fsp: SPP(s), fse | (FORALL (i : below[fse'length]) : member(fse(i), E)), fss) :
      fsp'length = fse'length & fsp'length = fss'length
      & (FORALL (i : below[fsp'length]) : subtermOF(s, fsp(i)) = ext(fss(i))(lhs(fse(i))))
      & t = replace_par_pos(s, fsp, fse, fss)
\end{verbatim}
}

{\tt RC(R)}, that is used in {\tt local\_joinability\_triangle?},
specifies the reflexive closure of the rewriting relation {\tt R}. For
a functional term {\tt s}, {\tt f(s)} and {\tt args(s)} compute the
head function symbol of {\tt s} and its arguments respectively; also,
{\tt app(f(s), args(s))} builds the functional term {\tt s}.  The
function {\tt replace} with arguments {\tt (t, st, i)} replaces the
{\tt (i+1)}$^{th}$ term of the sequence of arguments {\tt st} by {\tt
  t}.  The recursive function {\tt replaceTerm} replaces a subterm of
a term: it gives as output for the input triplet {\tt (s, t, p)} the
term obtained from {\tt s} by replacing the subterm at position {\tt
  p} of {\tt s} by {\tt t}, that in standard rewriting notation is
written as $s[p\leftarrow t]$.  Similarly, {\tt replace\_par\_pos}
specifies the parallel replacements necessary in one step of parallel
reduction. The specification of the relation of parallel reduction is
given by {\tt parallel\_reduction?}, in which the variables {\tt fsp,
  fse} and {\tt fss} are the sequences of parallel positions, rewrite
rules and substitutions, that were denoted respectively as $\Pi,
\Gamma$ and $\Sigma$, in the definition of the parallel reduction
relation.

The main lemmas and theorems specified and formalized about
orthogonality are presented below. All presented lemmas were
formalized. The lemma {\tt Linear\_and\_Non\_ambiguous\_implies\_}
{\tt confluent} is a weaker version of the lemma of confluence of
Orthogonal TRSs that is the last one.

{\footnotesize
\begin{verbatim}
- Linear_and_Non_ambiguous_implies_triangle: LEMMA FORALL (E) : 
  Linear?(E) & NOT Ambiguous?(E) => local_joinability_triangle?(reduction?(E))

- One_side_diamond_implies_conflent: LEMMA local_joinability_triangle?(R) => confluent?(R)   

- Linear_and_Non_ambiguous_implies_confluent: LEMMA
  FORALL (E) : ((Linear?(E) & NOT Ambiguous?(E) ) => confluent?(reduction?(E)))

- parallel_reduction: LEMMA 
  (reduction?(E)(t1, t2) => parallel_reduction?(E)(t1, t2)) 
   & (parallel_reduction?(E)(t1, t2) => RTC(reduction?(E))(t1, t2))

- parallel_reduction_is_DP: LEMMA Orthogonal?(E) => diamond_property?(parallel_reduction?(E))

- Orthogonal_implies_confluent: LEMMA 
   FORALL (E : Orthogonal) : LET RRE = reduction?(E) IN confluent?(RRE)
\end{verbatim}
}

{\tt RTC(R)}, that is used in {\tt parallel\_reduction}, specifies the
reflexive transitive closure of the rewriting relation {\tt R}.

The lemma {\tt Linear\_and\_Non\_ambiguous\_implies\_confluent} is
proved in a standard manner. In fact, since, in addition to
orthogonality restrictions, variables cannot appear repeatedly in the
rhs of the rules this proof does not need elaborated manipulation of
reductions and instantiations in order to build the term of parallel
joinability for divergence terms.

By the specification of these lemmas, one can observe that {\tt
  Orthogonal\_implies\_confluent}, that is the main lemma, depends on
the formalization of {\tt parallel\_reduction} and {\tt
  parallel\_reduction\_} {\tt is\_DP} . The former lemma is relatively
simple and the latter is the crucial one.

In order to classify overlaps in a parallel divergence from a term in
which, on the one side, a parallel reduction is applied at positions
$\Pi_1$ and, on the other side, at positions $\Pi_2$, positions
involved in a parallel divergence are classified through the following
specified recursive relations:

{\footnotesize
\begin{verbatim}
-sub_pos((fsp : PP), p : position): RECURSIVE finseq[position] =
     IF  length(fsp) = 0 THEN empty_seq[position]
     ELSIF p <= fsp(0) & p /= fsp(0) THEN add_first(fsp(0), sub_pos(rest(fsp), p))
        ELSE sub_pos(rest(fsp), p)
     ENDIF  MEASURE length(fsp)

-Pos_Over((fsp1 : PP), (fsp2 : PP)): RECURSIVE finseq[position] =
     IF length(fsp1) = 0 THEN empty_seq[position]
     ELSE (IF ( length(sub_pos(fsp2, fsp1(0))) > 0  
                 OR PP?(add_first(fsp1(0), fsp2)))
                THEN add_first(fsp1(0), Pos_Over(rest(fsp1), fsp2))
                ELSE Pos_Over(rest(fsp1), fsp2) ENDIF)
    ENDIF MEASURE length(fsp1)
\end{verbatim}
}

{\tt sub\_pos}$(\Pi,\pi)$ builds the subsequence of positions of the
sequence of parallel positions $\Pi$ that are strictly below the
position $\pi$; that is, $\pi'\in \Pi$ such that $\pi$ is a prefix of
$\pi'$, as usual denoted as $\pi<\pi'$.  {\tt
  Pos\_Over}$(\Pi_1,\Pi_2)$ builds the subsequence of positions from
$\Pi_1$ that are parallel to all positions in $\Pi_2$ or that have
positions in the sequence $\Pi_2$ below them. In this specification,
{\tt PP?} is a predicate for the type {\tt PP} of sequences of
parallel positions. These functions are crucial in order to build the
term of one-step parallel joinability, necessary in the proof of lemma
{\tt parallel\_reduction\_is\_DP}.

Confluence of orthogonal TRSs is proved according to the following
sketch: Firstly, it is proved
$\rightarrow\:\subseteq\:\rightrightarrows\:\subseteq\:\rightarrow^*$,
from which one concludes that
$\rightrightarrows^*\:=\:\rightarrow^*$. The lemma {\tt
  parallel\_reduction} corresponds to the latter inclusion.  Then, it
is proved that for orthogonal systems, $\rightrightarrows$ has the
diamond property, which corresponds to the lemma {\tt
  parallel\_reduction\_is\_DP}.  For an analytical proof see the
extended version of this paper.

\section{Formalization of Confluence of Non Ambiguous and Linear TRSs}

Computational formalizations do not admit mistakes and, in particular,
those specifications based on rewriting rules as well as on recursive
functional definitions can profit from a formalization of confluence
of orthogonality.  Several works report efforts on specification of
different computational objects (software and hardware) through TRSs
(e.g., \cite{ArSh1999b, MBARH2005, ARLaJaHa2006,MorraBCB08}).
Consequently, it is relevant to have robust and as complete as
possible libraries for the theory of abstract reduction systems and
TRSs in different proof assistants.  To construct the joinability term
for the lemma {\tt parallel\_reduction\_is\_DP}, one has to consider
several cases from which one is explained in the sequel.

Suppose, one has a parallel divergence from term $t_1\leftleftarrows s
\rightrightarrows t_2$ at positions $\Pi_1$ and $\Pi_2$ with
respective associated rules and substitutions $\Gamma_i$ and
$\Sigma_i$, $i\in\{1,2\}$. Let $\pi\in\Pi_2$ and ${\color{black}
  l\rightarrow r}$ and ${\color{black} \sigma}$ denote the associated
rule and substitution, respectively.  {\tt
  sub\_pos}$({\color{black}\Pi_1}, {\color{black} \pi})$ builds the
subsequence ${\color{black}\Pi_\pi}$ of positions in
${\color{black}\Pi_1}$ below ${\color{black}\pi}$.  Let
${\color{black}\Pi_{\pi}} = {\color{black} \{\pi_1, \ldots
  \pi_k\}}$. For $1\leq j\leq k$, let ${\color{black} g_j\rightarrow
  d_j}$ and ${\color{black} \sigma_j}$ denote the rule and
substitution associated with position ${\color{black} \pi_j}$. Then,
by non-ambiguity, for all $1\leq j \leq k$, there exist $\pi'_j$ and
$\pi''_j$ such that ${\color{black}\pi}\pi'_j\pi''_j =
{\color{black}\pi_j}$, being $\pi'_j$ a variable position of the lhs
of the rule ${\color{black} l\rightarrow r}$.

Let $\sigma'$ be the substitution obtained from ${\color{black}\sigma}$
modifying all variables according to substitutions
${\color{black}\sigma_j}$, then, the divergence at position
${\color{black}\pi}$, that is $t_2|_{\color{black}\pi} \leftleftarrows
s|_{\color{black}\pi} \rightrightarrows t_1|_{\color{black}\pi}$ can
be joined in one step of parallel reduction as
$t_2|_{\color{black}\pi} = {\color{black}r\sigma} \rightrightarrows
{\color{black}r}\sigma' \leftarrow {\color{black}l}\sigma'=
t_1|_{\color{black}\pi}$.  The construction of $\sigma'$ is one of the
most elaborated steps in this formalization.  Namely, suppose $x$ is a
variable occurring in the lhs of the rule ${\color{black} l\rightarrow
  r}$ only at position $\pi'$ (left-linearity guarantees unicity of
$\pi'$); if $\pi' \not= \pi_j'$, for all $1\leq j\leq k$, then
$x\sigma' := x{\color{black}\sigma}$. Otherwise, let $\{j_1, \ldots,
j_m\}$ be the set of indices such that $\pi' = \pi'_{j_l}$, for $1\leq
l \leq m$ and $1\leq j_l\leq k$. Since ${\color{black}\Pi_\pi}$ are
parallel positions, $\{\pi''_{j_1}, \ldots \pi''_{j_m}\}$ are parallel
positions of $x{\color{black}\sigma}$. By applying the rules
${\color{black}g_{j_l}\rightarrow d_{j_l}}$ with substitutions
${\color{black}\sigma_{j_l}}$, for $1\leq l \leq m$, one reduces in
parallel $x{\color{black}\sigma} \rightrightarrows
x{\color{black}\sigma} [\pi''_{j_1} \leftarrow
{\color{black}d_{j_1}\sigma_{j_1}}]\ldots[\pi''_{j_m} \leftarrow
{\color{black}d_{j_m}\sigma_{j_m}}]$. Thus, in this case, $x\sigma'$
is defined as $x{\color{black}\sigma} [\pi''_{j_1} \leftarrow
{\color{black}d_{j_1}\sigma_{j_1}}]\ldots[\pi''_{j_m} \leftarrow
{\color{black}d_{j_m}\sigma_{j_m}}]$.
   
The polymorphic function {\tt choose\_seq} below was specified to
construct associated subsequences of positions, rules or
substitutions.  {\tt choose\_seq}$({\color{black}\Pi_\pi},
{\color{black}\Pi_1}, {\color{black} \Gamma_1})$ and {\tt
  choose\_seq}$({\color{black}\Pi_\pi}, {\color{black}\Pi_1},
{\color{black} \Sigma_1})$ give respectively the subsequences of rules
and substitutions associated with ${\color{black}\Pi_\pi}$.  In
particular, {\tt choose\_seq} can be used in order to choose the
sequence of terms, instantiations of rhs's of rules, that should be
changed in order to obtain $x\sigma'$, for a variable $x$ occurring at
position ${\color{black}\pi}\pi'$. Namely, this is done calling {\tt
  choose\_seq(sub\_pos(${\color{black}\Pi_1},
  {\color{black}\pi}\pi'$), ${\color{black}\Pi_1},
  \{{\color{black}d_1\sigma_1}, \ldots,
  {\color{black}d_n\sigma_n}\}$)}, where the sequence of terms
$\{{\color{black}d_1\sigma_1}, \ldots, {\color{black}d_n\sigma_n}\}$
is straightforwardly built from the sequences of rules and
substitutions associated with ${\color{black}\Pi_1}$, i.e.,
${\color{black} \Gamma_1} = \{ {\color{black}g_1\rightarrow d_1},
\ldots, {\color{black}g_n\rightarrow d_n}\}$ and ${\color{black}
  \Sigma_1}=
\{{\color{black}\sigma_1},\ldots,{\color{black}\sigma_n}\}$.

{\footnotesize
\begin{verbatim}
 choose_seq(seq:PP, seq1:PP, (seq2 | seq1'length=seq2'length)): RECURSIVE finseq[T] = 
   IF length(seq)=0  THEN empty_seq
      ELSIF index(seq1,seq(0)) < seq1'length 
            THEN add_first(seq2(index(seq1,seq(0))), choose_seq(rest(seq),seq1,seq2))
            ELSE choose_seq(rest(seq),seq1,seq2)
   ENDIF  MEASURE(length(seq))
\end{verbatim}
}

The function {\tt index}$(\Pi,\pi)$ above returns the index of the
position $\pi$ in the sequence $\Pi$, which is less than the length of
$\Pi$, if $\pi$ occurs indeed in $\Pi$. Otherwise, it returns the length of $\Pi$.

The construction of $\sigma'$ requires the specification of two
recursive functions {\tt SIGMA} and {\tt SIGMAP}.

{\footnotesize
\begin{verbatim}
 SIGMA(sigma, x, fst, (fsp:SPP(sigma(x))|length(fsp)=length(fst)))(y:(V)): term = 
   IF length(fst)=0 OR y/=x 
      THEN sigma(y) 
      ELSE replace_terms(sigma(x),fst,fsp) 
   ENDIF
\end{verbatim}
}

{\tt SIGMA} has as arguments $\sigma$, $x$ and the associated
subsequences of substituting terms and positions relative to the
necessary update of $x\sigma$. One has, {\tt
  SIGMA}$({\color{black}\sigma}, x, \{{\color{black}
  d_{j_1}\sigma_{j_1}},\ldots , {\color{black}d_{j_m}\sigma_{j_m}}\},
\{ \pi_{j_1}'',\ldots, \pi_{j_m}''\})$ applied to $x$ will give
$x\sigma'$, that is $x{\color{black}\sigma} [\pi''_{j_1} \leftarrow
{\color{black}d_{j_1}\sigma_{j_1}}]\ldots[\pi''_{j_m} \leftarrow
{\color{black}r_{j_m}\sigma_{j_m}}]$.

The construction of the whole substitution $\sigma'$, is done through
the function {\tt SIGMAP} below, that adequately calls the function
{\tt SIGMA}.  {\tt SIGMAP}$({\color{black}\sigma}, \{x_1,\ldots,
x_q\}, \{ {\color{black}\pi}\pi'_1,\ldots, {\color{black}\pi}\pi'_q\},
\{{\color{black} d_{1}\sigma_{1}},\ldots ,
{\color{black}d_{n}\sigma_{n}}\}, \{ {\color{black}\pi_{1}},\ldots,
{\color{black}\pi_{n}}\})$, where $\{x_1,\ldots, x_q\}$ and $\{
{\color{black}\pi}\pi'_1,\ldots, {\color{black}\pi}\pi'_q\}$ are the
sequence of variables at lhs of the rule ${\color{black}l\rightarrow
  r}$ that should change, assuming ${\color{black}l\sigma}$ occurs at
position ${\color{black}\pi}$, and the associated sequence of
positions of these variables in the whole term $t_1$,
respectively. For a variable $y\in \{x_1,\ldots, x_q\}$, say $y=x_r$,
{\tt SIGMAP} calls the function {\tt SIGMA} giving as input the
sequence of terms to be substituted and their associated positions in
$y{\color{black}\sigma}$. This is done through application of the
functions {\tt choose\_seq} and {\tt complement\_pos}.  The former
one, is called as {\tt choose\_seq}$(
\{{\color{black}\pi}\pi'_r\pi''_{r,j_1},\ldots,{\color{black}\pi}\pi'_r\pi''_{r,j_{m_r}}\},
\{ {\color{black}\pi_{1}},\ldots, {\color{black}\pi_{n}}\},
\{{\color{black} d_{1}\sigma_{1}},\ldots ,
{\color{black}d_{n}\sigma_{n}}\})$, which gives the sequence of
substituting terms. The latter one is called as {\tt
  complement\_pos}$({\color{black}\pi}\pi'_r, \{
{\color{black}\pi_{1}},\ldots, {\color{black}\pi_{n}}\})$, which gives
as result the associated positions inside ${\color{black}l\sigma}$,
that is $\{\pi''_{r,j_1},\ldots,\pi''_{r,j_{m_r}}\}$.

{\footnotesize
\begin{verbatim}
 SIGMAP(sigma,fsv,(fsp1:PP|fsp1`length=fsv`length),
        fst,(fsp2:PP|fsp2`length=fst`length))(y:(V)): RECURSIVE term= 
   IF length(fsv)=0 
      THEN sigma(y)
      ELSIF y=fsv`seq(0) & SP?(sigma(fsv`seq(0)))(complement_pos(fsp1`seq(0),fsp2))
            THEN SIGMA(sigma,fsv`seq(0),choose_seq(sub_pos(fsp2,fsp1`seq(0)),fsp2,fst),
                             complement_pos(fsp1`seq(0),fsp2))(y)
            ELSE SIGMAP(sigma,rest(fsv),rest(fsp1),fst,fsp2)(y)
   ENDIF MEASURE(length(fsv))
\end{verbatim}
}

A small number of lemmas were formalized in order to prove soundness of
this definition. Namely, the fact that it is in fact a substitution is
axiomatized.  Among these lemmas, as a matter of illustration, it is
necessary to prove that the subsequences of terms and positions given
as third and second parameters of the call of {\tt SIGMA} have the
same length.

This is stated as the following lemma easily formalized by induction
on the length of the finite sequences. In fact, this lemma says that,
if one compares a position {\tt p} with a sequence of parallel
positions {\tt fsp}, the complementary positions are obtained from the
same positions that are under {\tt p}.

{\footnotesize
\begin{verbatim}
 complement_pos_preserv_sub_pos_length1: LEMMA
   PP?(fsp) => complement_pos(p, fsp)`length = sub_pos(fsp, p)`length
\end{verbatim}
}

Currently, the whole PVS {\tt orthogonal} development consist of among
1.300 lines of specification and 46.000 lines of proofs.  Indeed,
there are 40 definitions, 84 proved lemmas and 8 axioms.

\section{Related work and Conclusions}

PVS specifications of non trivial notions and formalizations of
results of the theory of term rewriting systems were presented, that are related
with the properties of the parallel rewriting reduction and orthogonal rewriting
systems.  The PVS \emph{theory} for orthogonal TRSs enriches the PVS
\emph{theory} {\tt trs} for TRSs introduced in \cite{GaAR2008b} and
available in \cite{NASAPVSLib}.  The formalization of these properties
of orthogonal TRSs are close to the analytical inductive proofs
presented in textbooks such as \cite{BaNi98} and \cite{Te2003} that in
essence are based in the well-known parallel moves lemma which projects
parallel reductions over a simple rewriting reduction. These
formalizations provide additional evidence of the appropriateness of
both the higher-order specification language and the proof engine of
PVS to deal in a natural way with specification of rewriting notions
and properties and their formalizations.  This consequently implies
the good support of PVS to deal with soundness and completeness and
integrity constraints of specifications of computational objects
specified through rewriting rules.

In its current status, the theory for orthogonal TRSs includes a
complete formalization of confluence of non-ambiguous and linear TRSs
as well as a proof of confluence of orthogonal TRSs by using standard
definitions and proof ideas shown in text books that ease the
understanding of them. The last theorem depends on both the lemma of
equivalence of the reflexive-transitive closure of the rewriting and
the parallel reduction relations and of the lemma of diamond property
of the parallel reduction relation of orthogonal TRSs.  The latter
lemma is formalized axiomatizing some technical properties of parallel
positions, rules and substitutions involved in one-step of parallel
reduction.  In \cite{Th2012} the criterion of weak orthogonality was
integrated to ensure confluence applying the certification tool
CeTA. Unlike orthogonality, weak orthogonality allows for trivial
critical pairs. To the best of our knowledge any complete formalization of the
property of confluence of orthogonal TRSs is available in any proof
assistant.

\bibliographystyle{eptcs}


\end{document}